\documentstyle[epsfig]{aipproc}

\begin{document}
\title{Study of $e^+e^- \rightarrow H^+H^-$\\ at a 800~GeV Linear Collider}

\author{A.~Kiiskinen$^{\dagger}$, M.~Battaglia$^{*}$ and 
P.~P\"oyh\"onen$^{\dagger}$}
\address{$^{\dagger}$Helsinki Institute of Physics, FIN 00014 Helsinki 
Finland \\
$^*$CERN, CH 1211 Geneva 23 Switzerland\\
and Dept.\ of Physics, University of Helsinki, FIN 00014 Finland}

\maketitle

\begin{abstract}
The production and decay of heavy charged Higgs bosons at a 800~GeV $e^+e^-$ 
linear collider have been studied. The analysis of the $H^+H^- \rightarrow 
t\bar{b} \bar{t}b$, expected to be dominant in the MSSM, and $H^+H^- 
\rightarrow W^+h^0 W^-h^0$ decay modes leading to the same final state 
consisting of two $W$ bosons and four $b$ quarks, provides with a 
determination of the boson mass to 1~GeV/$c^2$ and of the production cross 
section with 10\% accuracy for 500~fb$^{-1}$ of data.
\end{abstract}
 
\section*{Introduction}

The study of the origin of electro-weak symmetry breaking and the exploration
of the Higgs sector is one of the main themes of the $e^+e^-$ linear collider 
(LC) physics programme. A distinctive feature of several extensions of the 
Standard Model (SM), such as its minimal supersymmetric extension (MSSM), is 
the introduction of an extra Higgs doublet resulting in five physical Higgs 
boson states, two of which are charged $H^{\pm}$.  
While the existence of an extended Higgs sector may be indirectly revealed by 
a detailed study of the production and decay properties of the lightest 
neutral Higgs boson $h^0$~\cite{hreview}, its direct manifestation will only 
come from the detection of pairs of heavier Higgs particles, such as
$e^+e^- \rightarrow H^+H^-$. We report the results of a study of the 
sensitivity of a 800~GeV $e^+e^-$ LC, such as TESLA~\cite{tesla}, for the 
detection of pair produced heavy charged Higgs bosons.
The $e^+e^- \rightarrow H^+H^-$ production cross section depends, at tree 
level, only on the boson mass, $M_{H^{\pm}}$, while the dominant $H^{\pm}$ 
decay mode depends on the model parameters that may modify the Higgs couplings
to gauge bosons and fermions. There have been some earlier studies of
$H^+H^-$ reconstruction at the LC for the case of a rather light boson,  
decaying predominantly into a $cs$ quark pair~\cite{hpmlc}. This study has 
considered the two decay processes $H^+H^- \rightarrow t\bar{b}\bar{t}b$, 
expected to be dominant in the MSSM with $0.98 < {\mathrm{BR}} < 0.82$  for
$2 < \tan \beta <50$, and $H^+H^- \rightarrow W^+h^0W^-h^0$, leading to the 
same final state consisting of two $W$ bosons and four $b$ quarks, for the 
cases of heavier bosons, with $M_{H^{\pm}}$ = 200 and 300~GeV/$c^2$ and 
$M_{h^0}$ = 115~GeV/$c^2$. 
  
By selecting hadronic $W$ decays, the complete reconstruction of the resulting
eight jet final states allows to determine the $H^{\pm}$ mass, profiting of the
intermediate $t$ ($h^0$) and $W$ mass constraints, the production cross section
and the decay branching fractions. Accurate values for the mass and the decay
branching ratios may contribute to fix some of the model parameters. Recent 
calculations of higher order corrections to the production cross section
have shown a significant model dependence~\cite{rad_corr1} that may be useful 
for discriminating between models, provided an accurate experimental
determination of the cross section is possible at the LC.
  
The results reported here have been obtained from a simulation study for 
500~$fb^{-1}$ of integrated luminosity at $\sqrt{s}$ = 800~GeV. The signal and
backgrounds events have been generated with PYTHIA~6~\cite{pythia}, including 
initial state radiation and accounting for the beamstrahlung effect computed 
for the TESLA parameters~\cite{circe}. The generated events have been passed 
through the parametric SIMDET detector simulation~\cite{simdet}.

\section*{Event Selection}

The signal cross section depends strongly on the charged Higgs boson 
mass, varying from $\sim 29$~$fb$ for 200~GeV/$c^2$ to $\sim 12$~$fb$ for 
300~GeV/$c^2$ corresponding to $1.5 \times 10^4$--$6 \times 10^3$ $H^+H^-$ 
pairs/500 fb$^{-1}$. This abundant production at a high luminosity LC, allows 
to enforce tight requirements on the event preselection and the mass
reconstruction.
Most two and four fermion background processes can be effectively rejected
by requiring eight jets and four $b$ quarks, despite their large production
cross sections. The $t\bar{t}$ background which results in multi-jet final 
state with $b$ quarks can be reduced by the both $b$ tagging and the mass 
kinematical fits.  
Genuine $t\bar{b} \bar{t}b$ final states, as those originating from the Higgs 
radiation off the top quark process $e^+e^- \rightarrow t \bar{t} h^0 
\rightarrow t \bar{t} b \bar{b}$, remain the largest source of background with
an estimated cross section of 3~fb which has been reduced in this analysis 
by a kinematical fit. 

To select fully hadronic events, the visible energy of the reconstructed
tracks and calorimetric clusters has been required to be above 600~GeV, the 
missing momentum below 100~GeV/$c$ and the energy of the most energetic lepton
below 50~GeV. The hadronic system has been clustered into jets, using the 
CAMJET algorithm~\cite{camjet}, varying the $y_{cut}$ value from 
$2.6 \times~10^{-4}$ to $5.7 \times~10^{-5}$. Only those events giving eight 
reconstructed jets, for a $y_{cut}$ value in this range, have been further 
considered. Four out of the eight jets have been required to be $b$ tagged. 
In this analysis a parametrized $b$ tagging response, corresponding to 90\% 
tagging efficiency with 5\% mis-identification probability, has been used.  

\section*{Mass reconstruction}

In order to efficiently distinguish the signal charged Higgs
production from the underlying backgrounds and to measure the boson
mass, it is important to obtain a clean Higgs signal in the mass
distribution of the multi-jet final states.  The two decay hypotheses
$H^+H^- \rightarrow t\bar{b} \bar{t}b$ and $H^+H^- \rightarrow W^+h^0
W^-h^0$ have been considered. Events fulfilling the above selection
criteria have been tested for the presence of two $W$ decaying into
non-$b$ tagged jets. Each $t$ quark has then been reconstructed from a
$W$ candidate paired with a $b$ tagged jet, in the case of the $H^+
H^- \rightarrow t\bar{b}\bar{t}b$ while for the case of $H^+H^-
\rightarrow W^+h^0 W^-h^0$, each $h^0 \rightarrow b \bar b$ decay has
been reconstructed from tagged $b \bar b$ pairs. To further reject
backgrounds as well as poorly reconstructed signal events, leading to
an inaccurate measurement of $M_{H^{\pm}}$, the compatibility of the
mass of the jet combination to that of $W^{\pm}$ and $h^0$ bosons and
the top quark within the measurement accuracy has been required. After
completing the jet assignment, a kinematical fit, imposing energy and
momentum conservation, the mass of intermediate states and equal Higgs
boson masses, has been applied. This fit improves the signal mass
reconstruction from by a factor of two for $M_{H^{\pm}}$ =
300~GeV/$c^2$.

%\begin{figure}[h!] 
%\centerline{\epsfig{file=tbtb2013.eps,width=6cm}\epsfig{file=tbtb2016.eps,
%width=6cm}}
%\vspace{10pt}
%\caption{Fitted charged Higgs boson mass for $H^+ H^- \rightarrow 
%t\bar{b}\bar{t}b$ with $M_{H^{\pm}}$ = 300~GeV/$c^2$ for different cuts 
%corresponding to 6\% (left plot ) and 2\% (right plot) efficiency. 
%The histograms are normalized to 500~$fb^{-1}$ integrated luminosity and 
%100\% branching ratio to $t\bar{b}\bar{t}b$}
%\label{fig_mass1}
%\end{figure}
\begin{figure}[hb!] 
\centerline{\epsfig{file=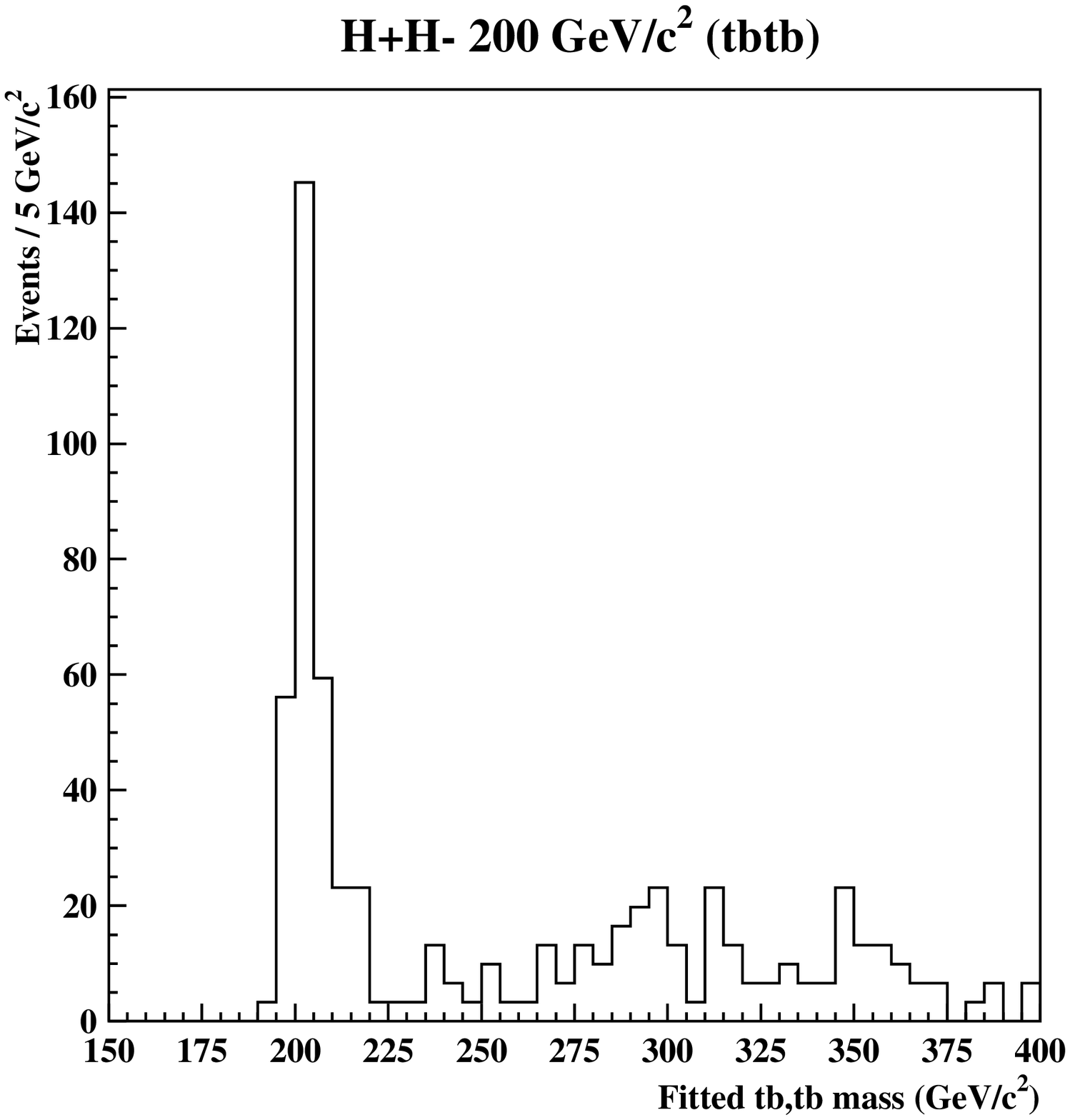,width=7cm}\epsfig{file=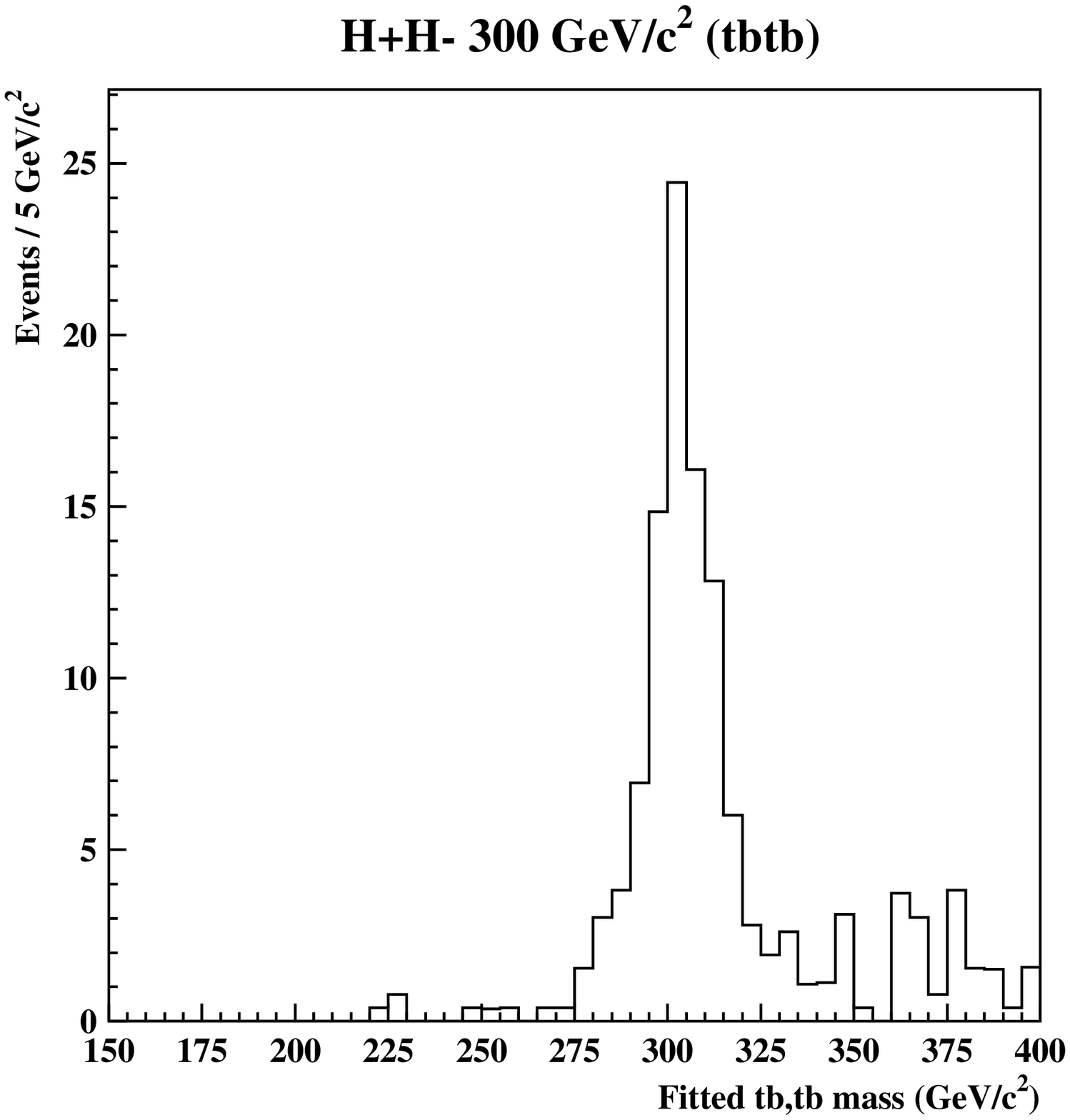,
width=7cm}}
\vspace{10pt}
\caption{Fitted charged Higgs boson mass for $H^+ H^- \rightarrow 
t\bar{b}\bar{t}b$ with $M_{H^{\pm}}$ = 200~GeV/$c^2$ (left plot) and 
$M_{H^{\pm}}$ = 300~GeV/$c^2$ (right plot). The histograms are normalized to 
500~$fb^{-1}$ integrated luminosity and 100\% branching ratio into the 
analyzed decay mode.}
\label{fig_mass2}
\end{figure}

The resulting fit $\chi^2$ has also been used to further reject background 
events and to distinguish between the two signal decay channels under study, 
based on the hypothesis giving the best $\chi^2$ value.

\section*{Results}

After applying these reconstruction and selection criteria, signal
efficiencies in the range 2\%-4\%, according to the charged Higgs
boson mass and adopted $\chi^2$ cut value, has been obtained with a
high signal purity. With a data set of 500~fb$^{-1}$ at $\sqrt{s}$ =
800~GeV, these correspond to signal samples of 150 - 600 events, with
$s/\sqrt{b}>100$. The corresponding statistical accuracies on the
measurements of the boson mass and the production cross section are
summarized in Table~1.

\begin{table}[h!]
\begin{center}
\caption{Statistical accuracy on the determination of the mass and production 
cross section for a charged Higgs boson at $\sqrt{s}$ = 800~GeV with
500~fb$^{-1}$}
\begin{tabular}{|c|c|c|}
\hline
$M_{H^{\pm}}$ (GeV/$c^2$) & $\delta$ $M_{H^{\pm}}$ (GeV/$c^2$) & 
$\delta \sigma_{H^+H^-} / \sigma_{H^+H^-}$ \\
\hline \hline
 200 & 0.4 & 0.06 \\
 300 & 1.1 & 0.10 \\ \hline
\end{tabular}
\end{center}
\end{table}

Due to the low level of background achieved, the sensitivity to a charged 
Higgs pair production is guaranteed up to $M_{H^{\pm}} \simeq$ 350~GeV/$c^2$, 
where the cross section falls below 5~fb, due to the $\beta$ suppression,
reducing the number of signal events below the detection threshold at a
$\sqrt{s}$ = 800~GeV collider.

\section*{Conclusions}

The full reconstruction of the charged Higgs boson $H^+H^- \rightarrow 
t\bar{b} \bar{t}b$ and $W^+h^0 W^-h^0$ multi-jet final states has been studied
for a sample of $e^+e^-$ LC data at $\sqrt{s}$ = 800~GeV with $M_{H^{\pm}}$ = 
200 and 300~GeV/$c^2$. Statistical accuracies of 1~GeV/$c^2$ on the boson 
mass and of 10\% on its production cross sections have been obtained. These
results can provide significant information about the parameters of an 
extended Higgs sector in the MSSM or in other SM extensions.


\begin{references}

\bibitem{hreview}
M.~Battaglia and K.~Desch, these proceedings.

\bibitem{tesla}
O.~Napoly, these proceedings.

\bibitem{hpmlc}
S.~Komamiya, Phys.\ Rev.\ {\bf D38} (1988) 2158.\\ 
P.~Eerola and J.~Sirkka, DESY 92-123A (1992) 133.\\
A.~Sopczak, Z.\ Phys.\ {\bf C65} (1995) 449.

\bibitem{rad_corr1}
J.~Guasch, W.~Hollik and A.~Kraft, hep-ph/9911452.\\
A.~Arhrib and G.~Moultaka, Nucl.\ Phys.\ {\bf B558} (1999) 3.

\bibitem{pythia} T.~Sj\"ostrand, Comp.\ Phys.\ Comm.\ {\bf 39}
  (1986), 347. Version 6.125 was used.

\bibitem{circe}
T.~Ohl, Comput.\ Phys.\ Commun.\ {\bf 101} (1997) 269

\bibitem{simdet}
H.J.~Schreiber, these proceedings.

\bibitem{camjet}
Y.~Dokshitzer, G.~Leder, S.~Moretti and B.~Webber, JHEP {\bf 08} (1997) 001

\end{references}
\end{document}